\documentclass[floatfix,aps,twocolumn,prl,superscriptaddress, amsfonts]{revtex4}

\usepackage{graphicx}

\usepackage{dcolumn}

\usepackage{bm}

\begin{document}
\title{Coupling Between An Optical Phonon and the Kondo Effect}
\author{K.S. Burch}
\email{kburch@lanl.gov}
\affiliation{Los Alamos National Laboratory, MS  K771, MPA-CINT, Los Alamos, NM 87545}
\author{Elbert E.M. Chia}
\affiliation{Los Alamos National Laboratory, MS  K771, MPA-CINT, Los Alamos, NM 87545}
\author{D. Talbayev}
\affiliation{Los Alamos National Laboratory, MS  K771, MPA-CINT, Los Alamos, NM 87545}
\author{B.C. Sales}
\affiliation{Condensed Matter Sciences Division, Oak Ridge National Laboratory, Oak Ridge, Tennessee 37831}
\author{D. Mandrus}
\affiliation{Condensed Matter Sciences Division, Oak Ridge National Laboratory, Oak Ridge, Tennessee 37831}
\author{A.J. Taylor}
\affiliation{Los Alamos National Laboratory, MS  K771, MPA-CINT, Los Alamos, NM 87545}
\author{R.D. Averitt}
\altaffiliation[Permanent address: ]
{Department of Physics, Boston Univeristy, 590 Commonwealth Avenue, Boston, MA 02215}
\affiliation{Los Alamos National Laboratory, MS  K771, MPA-CINT, Los Alamos, NM 87545}

\begin{abstract}
	We explore the ultra-fast optical response of Yb$_{14}$MnSb$_{11}$, providing further evidence that this Zintl compound is the first ferromagnetic, under-screened Kondo lattice. These experiments also provide the first demonstration of coupling between an optical phonon mode and the Kondo effect.\end{abstract}

\maketitle
For over forty years, there has been significant interest in the screening of local moments by conduction electrons, termed the Kondo effect. Since its discovery,\cite{KONDOJ:Resmdm}  the Kondo effect has produced a number of surprising results, including heavy Fermion (HF) compounds where the carriers are extremely massive.\cite{FISKZ:HEAMNH}  Kondo later proposed that the conduction electrons could also screen the motion of atoms,\cite{KondoJ:Locasm} an effect that could also result in exotic behavior.\cite{MitsumotoK:Phosad,YotsuhashiS:Newaqp,hotta:197201,NayakP:Eleiap} Despite extensive work on HF materials, there have been no observations of coupling between the Kondo effect and an optical phonon mode. 

The lattice plays an important role in the 14-1-11 HF compounds, where numerous properties can be altered through iso-electronic substitutions. The 14-1-11 compounds exhibit numerous ground-states, including magnetic order, metallic or semiconducting transport,\cite{growth,transport} and HF behavior\cite{sales_transport_yms,ymsoptics}. Of particular interest is Yb$_{14}$MnSb$_{11}$, which is metallic and exhibits ferromagnetic order of the Mn d$^{4}$ local moment below a Curie temperature (T$_{C}$) of 53 K. A study of its optical conductivity suggested that the Mn d-shell has a fifth electron that is screened by the Kondo effect. It was argued that the coexistence of ferromagnetism and HF behavior occurs via a distortion of the MnSb$_{4}$ tetrahedron;\cite{ymsoptics} motivating us to search for Kondo-lattice coupling in Yb$_{14}$MnSb$_{11}$. 

We demonstrate Kondo-phonon coupling in Yb$_{14}$MnSb$_{11}$ by simultaneously monitoring the electronic structure and phonon dynamics via ultra-fast optical experiments.\cite{DemsarHF, MerlinReview} Specifically, the relaxation time and amplitude of the photo-induced response strongly increase at low temperatures (T), indicating the presence of a hybridization gap in Yb$_{14}$MnSb$_{11}$. Furthermore, a softening of a phonon at low T, is correlated with the development of this HF state. 

The Yb$_{14}$MnSb$_{11}$ single crystals were grown as described elsewhere.\cite{transport, growth} This compound has been well characterized by X-ray diffraction,  magnetic susceptibility,\cite{growth} electrical resistivity, optical conductivity\cite{ymsoptics} specific-heat \cite{transport} and X-Ray Magneto Circular Dichroism\cite{xmcd}. The sample in this study provided mirror like surfaces and therefore time-domain measurements were performed on the as-grown sample. The ultra-fast experiments were conducted with a cavity dumped laser producing $50~fs$ pulse centered at $800~nm$ with a $2~MHz$ repetition rate. Comparable results were obtained with a 250 KHz regenerative amplifier ($800~nm$, $100~fs$) coupled to a superconducting magnet, which was used to investigate the effects of magnetic field and fluence. The pump-pulse was kept at a fluence of $\approx50~\frac{\mu J}{cm^{2}}$. 

\begin{figure*}
\includegraphics{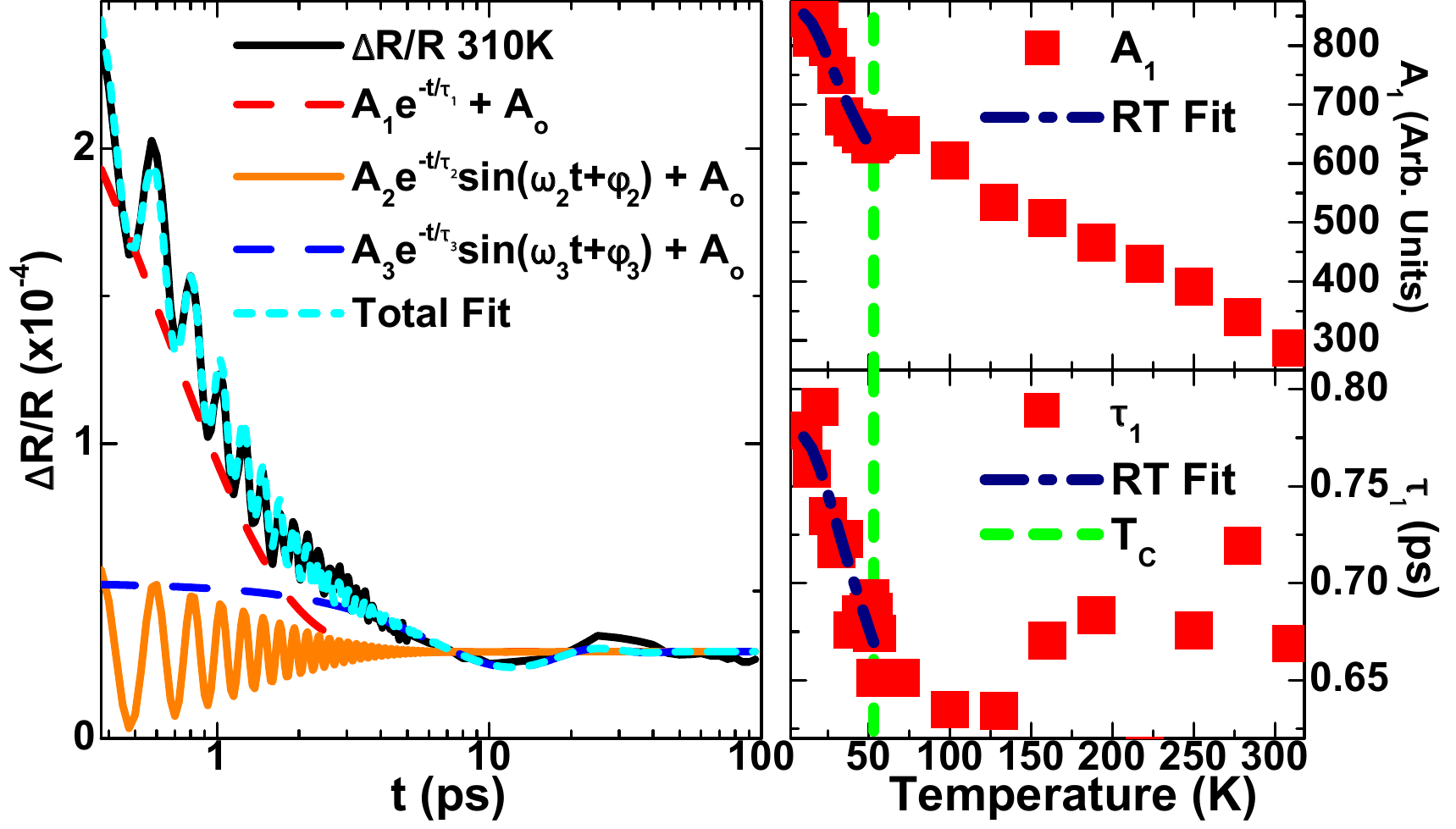}
\caption{\label{fig:dRoverR} Left: Pump induced change in reflectivity versus time. Data (black line) are fit with three terms, one exponential (red dashed line) and two exponentially damped sinusoidal signals (orange and dashed blue lines). Right: Amplitude (top) and relaxation time (bottom) of the exponential term versus temperature. Dramatic increases in A and $\tau$ at low temperatures are described via a gap ($\Delta=32~cm^{-1}$) in the density of states for $k_{B}T<\Delta$ by the Rothwarf-Taylor model (Navy dashed line).}
\end{figure*}

In the left panel of Fig. \ref{fig:dRoverR} we plot the photo-induced change in the reflectivity ($\frac{\Delta R(t)}{R}$) as function of time with the sample at 310 K. After the pump pulse arrives an increase in $\frac{\Delta R(t)}{R}$ indicates a rise in the temperature of the Fermi surface\cite{ymsoptics} as electron-electron interactions bring the optically excited carriers into equilibrium.\cite{groeneveld} We then observe an exponential decay and two exponentially damped oscillatory signals. The exponential decay can be attributed to the carrier and phonon baths coming into thermal equilibrium.\cite{groeneveld}

	To quantitatively evaluate our results we have fit the data at all temperatures to the formula: 
\begin{equation}
\label{eq:fit}
\frac{\Delta R(t)}{R}=A_{1}e^{-t/\tau_{1}}+\sum_{i=2,3}A_{i}e^{-t/\tau_{i}}\sin(\omega_{i}t+\phi_{i})+A_{0}
\end{equation}
The resulting $A_{1}$ and $\tau_{1}$ are shown in the right panels of  Fig. \ref{fig:dRoverR}. For $T>T_{C}$, $A_{1}$ and $\tau_{1}$ follow our expectations for a normal metal. Specifically $\tau_{1}$ is mostly temperature independent, while the smaller relative change in T after photo-excitation results in a reduction of $A_{1}$ as the temperature is raised.\cite{groeneveld} However, the decay time and amplitude increase significantly below $T_{C}$. Similar enhancements in the $\tau$ and A at low temperatures have been observed in  superconductors\cite{AverittYBCO,KaindlRA:UltmrY,PhysRevB.59.1497} and HF compounds.\cite{DemsarHF} This low T increase is believed to indicate the opening of a gap ($\Delta$) in the density of states. Specifically, due to the presence of a gap, the carriers can only relax via emission of high frequency phonons with energy $\hbar\omega_{phonon}\geq\Delta$. Since the probability for optical phonons to decay into two phonons of lower energy is fairly small (as opposed to acoustic modes that are much more susceptible to anharmonic effects), it is highly likely that the optical phonon will excite a new electron-hole pair across the gap. This \textit{phonon bottleneck} significantly reduces the transmission of heat from the Fermi surface to the lattice, resulting in an increase in $\tau$ and A when $k_{B}T<\Delta$. 	

	The model of Rothwarf and Taylor (RT) has been very successful in describing the dynamics of superconductors and HF compounds.\cite{RToriginal} In the RT approach the quasi-particle and phonon dynamics are  described via a set of coupled nonlinear differential equations. The RT equations account for the creation and destruction of electron hole pairs via emission of  high frequency phonons and bi-particle scattering as well as a reduction in the number of high frequency phonons via diffusion and anharmonic decay. Recently the RT model has been solved in a variety of cases leading to the solution:\cite{RTsolution} 
\begin{equation}
\label{eq:RTA}
A(T)=\frac{A_{0}}{n_{T}(T)+1}
\end{equation} 
\begin{equation}
\label{eq:RTtau}
\tau(T)=\tau_{0}(\delta(\epsilon n_{T}(T)+1)^{-1}+2n_{T}(T))^{-1}
\end{equation}
	where $n_{T}(T)$ is the population of the thermally excited quasi-particles. The results of simultaneously fitting $A_{1}(T)$ and $\tau_{1}(T)$ with eqs. \ref{eq:RTA} \& \ref{eq:RTtau} are displayed in the right panels of Fig. \ref{fig:dRoverR}. In performing these fits we used a standard form for the thermally excited quasi-particle density: $n_{T}(T)=(\Delta T)^{p}exp(\Delta/k_{B}T)$, where $0<p<1$ depends on the shape of the density of states. Similar to what has been observed in many HF compounds\cite{DemsarHF}, we found $p=0.49$ and a temperature independent gap of $\Delta=32~cm^{-1}$.  Interestingly, previous studies of the optical conductivity have found $\Delta\approx50~cm^{-1}$.\cite{ymsoptics} However optical conducitivty probes the direct hybridization gap (ie: no change in momentum), whereas the present results are governed by the indirect gap, which should be smaller.\cite{Coleman,Cox,Millis} Lastly, the application of a 6 T magnetic field, saturating the magnetization (M(T)), did not effect $\frac{\Delta R(t)}{R}$ at either 10 K or T$_{C}$, providing further evidence that the gap is unrelated to the ferromagnetism. 

\begin{figure}
\includegraphics{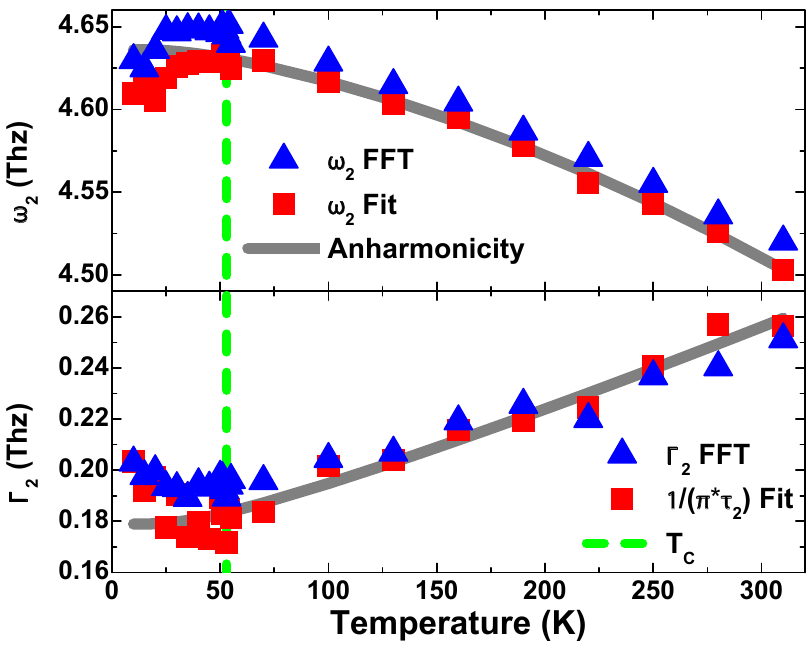}
\caption{\label{fig:params2} (color online) The phonon mode frequency (top panel) and broadening/damping rate (bottom panel). Red squares are results determined via eq. \ref{eq:fit}, and blue triangles are the result of fitting the FFT in Fig \ref{fig:freq} with a lorentzian line shape. The mode hardens and sharpens as the temperature is lowered to T$_{C}$, which is described by anharmonic effects (eqs. \ref{eq:anharmFreq} \& \ref{eq:anharmGamma}).}
\end{figure}
	 
\begin{figure}
\includegraphics{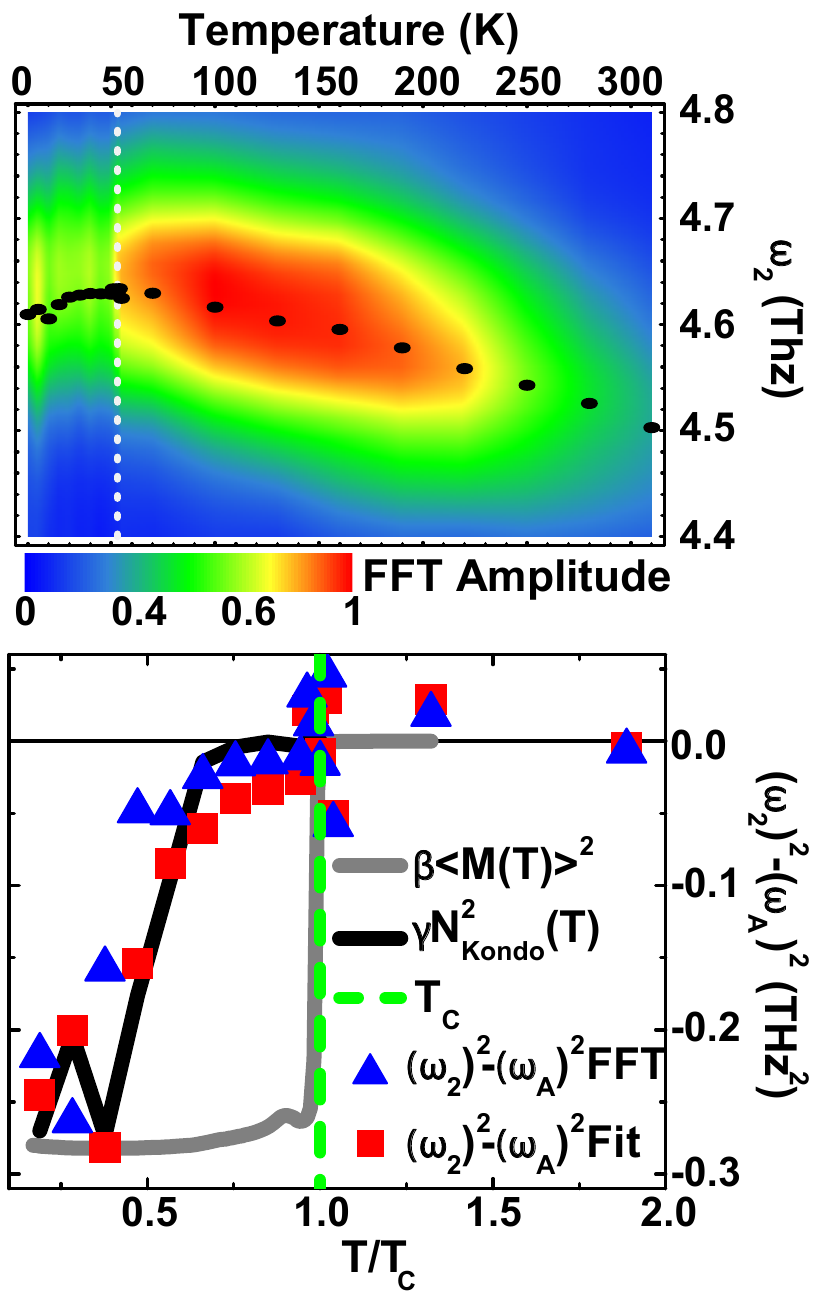}
\caption{\label{fig:freq} (color online) Top panel: FFT of $\frac{\Delta R(t)}{R} - A_{1}e^{-t/\tau_{1}}$, white dashed line indicates $T_{C}$ and black dots are the $\omega_{0}$. Bottom panel: The shift in the phonon frequency after anharmonic effects have been removed. The shift does not appear to result from coupling to ferromagnetic state (grey), but rather via a coupling to the Kondo effect (Black line).}
\end{figure}

	So far we have focused on the relaxation dynamics of the electronic system. However the data displayed in Fig. \ref{fig:dRoverR} also contains information about an optical phonon mode. Specifically, the oscillations in the data result from the coherent generation of phonons. While there are a variety of proposed mechanisms, it is generally agreed believed that such oscillations are initiated via a Raman process upon pump-pulse excitation. The optical mode is then observed via a Raman shift of the probe pulse.\cite{MerlinReview} The oscillatory signal exponentially decays due to the de-phasing of the optical modes. As shown in Fig. \ref{fig:params2},  as the temperature is lowered, the optical mode frequency ($\omega_{2}$) increases while its decay rate ($\frac{1}{\pi\tau_{2}}$) decreases. Whereas for $T < T_{C}$, the phonon softens and broadens. As discussed later, this low temperature behavior is quite anomalous. Therefore, to confirm that this behavior is intrinsic to the sample and not an artifact of our fitting routine, we subtracted the first term of eq. \ref{eq:fit} from the data and then performed a fast-Fourier transform (FFT). The FFT spectra in the top panel of Fig. \ref{fig:freq} reveal the same trends shown in Fig. \ref{fig:params2}. The center frequency and broadening determined from fits of the FFT data with Lorentzian line-shapes are in superb agreement with the results of fitting $\frac{\Delta R(t)}{R}$ (see Fig. \ref{fig:params2}). 

	The temperature dependence of the phonon above T$_{C}$ is similar to anharmonic effects observed in numerous other materials. Specifically, the optical phonon can decay into acoustic modes of lower energy, resulting in a re-normlization of its self-energy. Since the likelihood of this decay depends on the occupancy of the phonon density of states, one expects a T dependence of the form: 

\begin{eqnarray}
\label{eq:anharmGamma}
\Gamma_{A}(T) & = & \Gamma_{0}+a(1+2n_{B}(\frac{\omega_{0}}{2},T))+{}
\nonumber\\
&&{}b(1+3(n_{B}(\frac{\omega_{0}}{3},T)+n_{B}^{2}(\frac{\omega_{0}}{3},T)))\\
\label{eq:anharmFreq}
\omega_{A}(T) & = & \omega_{0} + c(1+2n_{B}(\frac{\omega_{0}}{2},T))+\nonumber\\
&&{} d(1+3(n_{B}(\frac{\omega_{0}}{3},T)+n_{B}^{2}(\frac{\omega_{0}}{3},T)))
\end{eqnarray} 

where $\omega_{0}$ is the intrinsic frequency from harmonic terms, $n_{B}(\omega,T)=\frac{1}{exp(\hbar\omega/k_{B}T)-1}$ is the Bose occupation function and we have assumed the phonon decays into two modes of frequency $\frac{\omega_{0}}{2}$ or three of frequency $\frac{\omega_{0}}{3}$.\cite{CardonaSiAnharm,HaroAnharmSi,LaxAnharmSi} As shown in Fig. \ref{fig:params2}, eqs. \ref{eq:anharmGamma} and \ref{eq:anharmFreq}  describe $\omega_{2}(T>T_{C})$ and $\Gamma_{2}(T>T_{C})$ well. Furthermore the values of $\frac{a}{\omega_{0}}$,$\frac{b}{\omega_{0}}$, $\frac{c}{\omega_{0}}$, $\frac{d}{\omega_{0}}$ are comparable to what has been observed in Si.\cite{CardonaSiAnharm,HaroAnharmSi,LaxAnharmSi} Nonetheless this does not explain the T-dependence of the phonon mode below $T_{C}$. 
\begin{table}

\caption{\label{TBL} Left: Anharmonic fit results for the optical phonon. Right: Fit parameters from Rothwarf-Taylor model.}
\begin{ruledtabular}
 
 \begin{tabular}{cccc}
$\omega_{0}$&
4.65 Thz&
A$_{0}$&
857
\tabularnewline
\hline
$\Gamma_{0}$& 
.17 Thz&
$\tau_{0}$&
0.006 Thz
\tabularnewline
\hline
$\frac{a}{\omega_{0}}$&
2.78$\times10^{-3}$&
$\delta$&
4.94
\tabularnewline
\hline
$\frac{b}{\omega_{0}}$&
3.85$\times10^{-5}$&
$\epsilon$&
0
\tabularnewline
\hline
$\frac{c}{\omega_{0}}$&
-3.49$\times10^{-3}$&
$\Delta$&
46 K (32 cm$^{-1}$)
\tabularnewline
\hline
$\frac{d}{\omega_{0}}$& 
-2.49$\times10^{-4}$&
p&
0.49
\tabularnewline
\end{tabular}

\end{ruledtabular}

\end{table}

	We now explore the origin of the $T<T_{C}$ behavior of the optical phonon, via the potential energy (U) for atomic displacements from equilibrium (x). The potential energy should have a form: $U=U_{0}+E_{FM}(x)<$M$>^2$, where $U_{0}$ includes the anharmonic and harmonic terms (ie: $\frac{d^{2}U_{0}}{dx^{2}}=\omega_{a}^{2}$), E$_{FM}(x)$ is the energy gained from the formation of the ferromagnetic state and $<$M(T)$>$ $=\frac{M(T)}{M(0)}$. Solving for the T dependence of the phonon frequency: 
	\begin{equation}
	\label{eq:phonon}
	\omega_{2}^{2} (T) = \frac{d^{2}U}{dx^{2}} =  \frac{d^{2}U_{0}}{dx^{2}} + \frac{d^{2}E_{FM}(x)}{dx^{2}}<M(T)>^2
	\end{equation}
Assuming $\frac{d^{2}E_{FM}(x)}{dx^{2}}$ is independent of x, we expect $\omega_{2}^{2} - \omega_{A}^{2}= \beta <$M(T)$>^{2}$.\cite{MATSUSHITAM:Anotdt,PhysRevB.28.1983,WAKAMURAK:Obssfe} However, in Fig. \ref{fig:freq} we observe that the phonon continues to soften down to the lowest T, whereas $\beta <$M(T)$>^{2}$ determined via zero-field cooled SQUID measurements, saturates close to T$_{C}$. 

	While we only considered the effect of ferromagnetic order on the phonon, the approach outlined above can be generalized to any interaction in the system whose strength depends on x. Since the binding energy of the Kondo singlets will depend on the hybridization between the Mn and Sb, we can replace E$_{FM}(x)$ by the Kondo binding energy (T$_{K}$) and $<$M$>$ by the relative number of singlets N$_{Kondo}(T)$. Next we estimate the number of Kondo singlets via the density of quasi-particles excited across the hybridization gap: $N_{Kondo} (T) = (1-\frac{n_{T} (T)}{n_{T} (T_{C})})$. Consistent with the Kondo-phonon coupling theory outlined above we find $\omega_{2}^{2} - \omega_{A}^{2}=\gamma N_{Kondo}^{2}(T)$ (see Fig. \ref{fig:freq}). 
		
	The fact that we observe strong coupling of the optical phonon to the formation of Kondo singlets and not to the ferromagnetic state, may not be surprising for three reasons. First, T$_{K}\approx300$ K\cite{sales_transport_yms,ymsoptics} is much bigger than T$_{C}=53$ K. Second, T$_{K}\propto\frac{1}{\rho}exp(-\frac{1}{J\rho})$ whereas E$_{FM}\propto J^{2}\rho$, where J is the exchange between the carriers and the local moments and $\rho$ is the density of states at the Fermi energy.\cite{Doniach_kondalattice} This implies that T$_{K}$ has greater sensitivity to atomic displacements than E$_{FM}$. Finally, we note that the scenario proposed in ref. \onlinecite{ymsoptics} for the under-screened Kondo effect in Yb$_{14}$MnSb$_{11}$, required a distortion of the MnSb$_{4}$ tetrahedron in order to break the degeneracy of the d$^{5}$ electrons. Therefore one would expect a coupling between the lattice and the Kondo effect.
	
	In conclusion, we explored the ultra-fast optical properties of Yb$_{14}$MnSb$_{11}$. Significant changes in the dynamics occur below $T_{C}$, although they are not connected with the development of the ferromagnetic state. In particular, the amplitude and relaxation time of the electronic response grows significantly at low temperatures, but is not effected by the application of a magnetic field. Furthermore, changes in the frequency of the optical phonon are inconsistent with the temperature dependence of the magnitization. Instead we find that the low temperature enhancement of the photo-induced response and the softening of the phonon result from the development of heavy Fermions and their coupling to the lattice. As expected for a hybridization gap, quantitative analysis reveals the indirect gap probed here is smaller in magnitude than the direct gap determined via optical conductivity experiments\cite{ymsoptics}. In addition, the softening of the phonon at low temperatures is correlated with the number of heavy quasi-particles. This is therefore the first demonstration of coupling between the Kondo effect and optical modes.
		
This work was supported by the DOE and NSF. Oak Ridge National Laboratory is managed by UT-Battelle, LLC, for the U.S. Dept. of Energy under contract DE-AC05-00OR22725. We are grateful for discussions with  S. Dordevic, I. Martin, R. Merlin, S. Trugman, and Jian-Xin Zhu.


\begin{thebibliography}{30}
\expandafter\ifx\csname natexlab\endcsname\relax\def\natexlab#1{#1}\fi
\expandafter\ifx\csname bibnamefont\endcsname\relax
  \def\bibnamefont#1{#1}\fi
\expandafter\ifx\csname bibfnamefont\endcsname\relax
  \def\bibfnamefont#1{#1}\fi
\expandafter\ifx\csname citenamefont\endcsname\relax
  \def\citenamefont#1{#1}\fi
\expandafter\ifx\csname url\endcsname\relax
  \def\url#1{\texttt{#1}}\fi
\expandafter\ifx\csname urlprefix\endcsname\relax\def\urlprefix{URL }\fi
\providecommand{\bibinfo}[2]{#2}
\providecommand{\eprint}[2][]{\url{#2}}

\bibitem[{\citenamefont{Kondo}(1964)}]{KONDOJ:Resmdm}
\bibinfo{author}{\bibfnamefont{J.}~\bibnamefont{Kondo}},
  \bibinfo{journal}{Progress of theoretical physics}
  \textbf{\bibinfo{volume}{32}}, \bibinfo{pages}{37 } (\bibinfo{year}{1964}).

\bibitem[{\citenamefont{Fisk et~al.}(1988)\citenamefont{Fisk, Hess, Petchick,
  Pines, Smith, Thompson, and Willis}}]{FISKZ:HEAMNH}
\bibinfo{author}{\bibfnamefont{Z.}~\bibnamefont{Fisk}},
  \bibinfo{author}{\bibfnamefont{D.~W.} \bibnamefont{Hess}},
  \bibinfo{author}{\bibfnamefont{C.~J.} \bibnamefont{Petchick}},
  \bibinfo{author}{\bibfnamefont{D.}~\bibnamefont{Pines}},
  \bibinfo{author}{\bibfnamefont{J.~L.} \bibnamefont{Smith}},
  \bibinfo{author}{\bibfnamefont{J.~D.} \bibnamefont{Thompson}},
  \bibnamefont{and} \bibinfo{author}{\bibfnamefont{J.~O.}
  \bibnamefont{Willis}}, \bibinfo{journal}{Science}
  \textbf{\bibinfo{volume}{239}}, \bibinfo{pages}{33 } (\bibinfo{year}{1988}).

\bibitem[{\citenamefont{Kondo}(1976)}]{KondoJ:Locasm}
\bibinfo{author}{\bibfnamefont{J.}~\bibnamefont{Kondo}},
  \bibinfo{journal}{Physica B + C} \textbf{\bibinfo{volume}{84B+C}},
  \bibinfo{pages}{40 } (\bibinfo{year}{1976}).

\bibitem[{\citenamefont{Mitsumoto and Ono}(2005)}]{MitsumotoK:Phosad}
\bibinfo{author}{\bibfnamefont{K.}~\bibnamefont{Mitsumoto}} \bibnamefont{and}
  \bibinfo{author}{\bibfnamefont{Y.}~\bibnamefont{Ono}},
  \bibinfo{journal}{Physica. C, Superconductivity}
  \textbf{\bibinfo{volume}{426-431}}, \bibinfo{pages}{330 }
  (\bibinfo{year}{2005}).

\bibitem[{\citenamefont{Yotsuhashi et~al.}(2005)\citenamefont{Yotsuhashi,
  Kojima, Kusunose, and Miyake}}]{YotsuhashiS:Newaqp}
\bibinfo{author}{\bibfnamefont{S.}~\bibnamefont{Yotsuhashi}},
  \bibinfo{author}{\bibfnamefont{M.}~\bibnamefont{Kojima}},
  \bibinfo{author}{\bibfnamefont{H.}~\bibnamefont{Kusunose}}, \bibnamefont{and}
  \bibinfo{author}{\bibfnamefont{K.}~\bibnamefont{Miyake}},
  \bibinfo{journal}{Journal of the Physical Society of Japan}
  \textbf{\bibinfo{volume}{74}}, \bibinfo{pages}{49 } (\bibinfo{year}{2005}).

\bibitem[{\citenamefont{Hotta}(2006)}]{hotta:197201}
\bibinfo{author}{\bibfnamefont{T.}~\bibnamefont{Hotta}},
  \bibinfo{journal}{Phys. Rev. Lett.} \textbf{\bibinfo{volume}{96}},
  \bibinfo{eid}{197201} (\bibinfo{year}{2006}).

\bibitem[{\citenamefont{Nayak et~al.}(2002)\citenamefont{Nayak, Ojha, Mohanty,
  and Behera}}]{NayakP:Eleiap}
\bibinfo{author}{\bibfnamefont{P.}~\bibnamefont{Nayak}},
  \bibinfo{author}{\bibfnamefont{B.}~\bibnamefont{Ojha}},
  \bibinfo{author}{\bibfnamefont{S.}~\bibnamefont{Mohanty}}, \bibnamefont{and}
  \bibinfo{author}{\bibfnamefont{S.~N.} \bibnamefont{Behera}},
  \bibinfo{journal}{IJMP B} \textbf{\bibinfo{volume}{16}}, \bibinfo{pages}{3595
  } (\bibinfo{year}{2002}).

\bibitem[{\citenamefont{Demsar et~al.}(2006)\citenamefont{Demsar, Sarrao, and
  Taylor}}]{DemsarHF}
\bibinfo{author}{\bibfnamefont{J.}~\bibnamefont{Demsar}},
  \bibinfo{author}{\bibfnamefont{J.~L.} \bibnamefont{Sarrao}},
  \bibnamefont{and} \bibinfo{author}{\bibfnamefont{A.~J.}
  \bibnamefont{Taylor}}, \bibinfo{journal}{J. Phys. Cond. Mat.}
  \textbf{\bibinfo{volume}{18}}, \bibinfo{pages}{R281} (\bibinfo{year}{2006}).

\bibitem[{\citenamefont{Merlin}(1997)}]{MerlinReview}
\bibinfo{author}{\bibfnamefont{R.}~\bibnamefont{Merlin}},
  \bibinfo{journal}{Solid State Communications} \textbf{\bibinfo{volume}{102}},
  \bibinfo{pages}{207 } (\bibinfo{year}{1997}).

\bibitem[{\citenamefont{Chan et~al.}(1998)\citenamefont{Chan, Olmstead,
  Kauzlarich, and Webb}}]{growth}
\bibinfo{author}{\bibfnamefont{J.}~\bibnamefont{Chan}},
  \bibinfo{author}{\bibfnamefont{M.}~\bibnamefont{Olmstead}},
  \bibinfo{author}{\bibfnamefont{S.}~\bibnamefont{Kauzlarich}},
  \bibnamefont{and} \bibinfo{author}{\bibfnamefont{D.}~\bibnamefont{Webb}},
  \bibinfo{journal}{Chem. Mat.} \textbf{\bibinfo{volume}{10}},
  \bibinfo{pages}{3583} (\bibinfo{year}{1998}).

\bibitem[{\citenamefont{Fisher et~al.}(1999)\citenamefont{Fisher, Wiener,
  Bud\char39{}ko, Canfield, Chan, and Kauzlarich}}]{transport}
\bibinfo{author}{\bibfnamefont{I.~R.} \bibnamefont{Fisher}},
  \bibinfo{author}{\bibfnamefont{T.~A.} \bibnamefont{Wiener}},
  \bibinfo{author}{\bibfnamefont{S.~L.} \bibnamefont{Bud\char39{}ko}},
  \bibinfo{author}{\bibfnamefont{P.~C.} \bibnamefont{Canfield}},
  \bibinfo{author}{\bibfnamefont{J.~Y.} \bibnamefont{Chan}}, \bibnamefont{and}
  \bibinfo{author}{\bibfnamefont{S.~M.} \bibnamefont{Kauzlarich}},
  \bibinfo{journal}{Phys. Rev. B} \textbf{\bibinfo{volume}{59}},
  \bibinfo{pages}{13829} (\bibinfo{year}{1999}).

\bibitem[{\citenamefont{Sales et~al.}(2005)\citenamefont{Sales, Khalifah, Enck,
  Nagler, Sykora, Jin, and Mandrus}}]{sales_transport_yms}
\bibinfo{author}{\bibfnamefont{B.~C.} \bibnamefont{Sales}},
  \bibinfo{author}{\bibfnamefont{P.}~\bibnamefont{Khalifah}},
  \bibinfo{author}{\bibfnamefont{T.~P.} \bibnamefont{Enck}},
  \bibinfo{author}{\bibfnamefont{E.~J.} \bibnamefont{Nagler}},
  \bibinfo{author}{\bibfnamefont{R.~E.} \bibnamefont{Sykora}},
  \bibinfo{author}{\bibfnamefont{R.}~\bibnamefont{Jin}}, \bibnamefont{and}
  \bibinfo{author}{\bibfnamefont{D.}~\bibnamefont{Mandrus}},
  \bibinfo{journal}{Phys. Rev. B} \textbf{\bibinfo{volume}{72}}
  (\bibinfo{year}{2005}).

\bibitem[{\citenamefont{Burch et~al.}(2005)\citenamefont{Burch, Schafgans,
  Butch, Sayles, Maple, Sales, Mandrus, and Basov}}]{ymsoptics}
\bibinfo{author}{\bibfnamefont{K.~S.} \bibnamefont{Burch}},
  \bibinfo{author}{\bibfnamefont{A.}~\bibnamefont{Schafgans}},
  \bibinfo{author}{\bibfnamefont{N.~P.} \bibnamefont{Butch}},
  \bibinfo{author}{\bibfnamefont{T.~A.} \bibnamefont{Sayles}},
  \bibinfo{author}{\bibfnamefont{M.~B.} \bibnamefont{Maple}},
  \bibinfo{author}{\bibfnamefont{B.~C.} \bibnamefont{Sales}},
  \bibinfo{author}{\bibfnamefont{D.}~\bibnamefont{Mandrus}}, \bibnamefont{and}
  \bibinfo{author}{\bibfnamefont{D.~N.} \bibnamefont{Basov}},
  \bibinfo{journal}{Phys. Rev. Lett.} \textbf{\bibinfo{volume}{95}},
  \bibinfo{pages}{046401} (\bibinfo{year}{2005}).

\bibitem[{\citenamefont{Holm et~al.}(2002)\citenamefont{Holm, Kauzlarich,
  Morton, Waddill, Pickett, and Tobin}}]{xmcd}
\bibinfo{author}{\bibfnamefont{A.}~\bibnamefont{Holm}},
  \bibinfo{author}{\bibfnamefont{S.}~\bibnamefont{Kauzlarich}},
  \bibinfo{author}{\bibfnamefont{S.}~\bibnamefont{Morton}},
  \bibinfo{author}{\bibfnamefont{G.}~\bibnamefont{Waddill}},
  \bibinfo{author}{\bibfnamefont{W.}~\bibnamefont{Pickett}}, \bibnamefont{and}
  \bibinfo{author}{\bibfnamefont{J.}~\bibnamefont{Tobin}},
  \bibinfo{journal}{JACS} \textbf{\bibinfo{volume}{124}}, \bibinfo{pages}{9894}
  (\bibinfo{year}{2002}).

\bibitem[{\citenamefont{Groeneveld et~al.}(1995)\citenamefont{Groeneveld,
  Sprik, and Lagendijk}}]{groeneveld}
\bibinfo{author}{\bibfnamefont{R.~H.~M.} \bibnamefont{Groeneveld}},
  \bibinfo{author}{\bibfnamefont{R.}~\bibnamefont{Sprik}}, \bibnamefont{and}
  \bibinfo{author}{\bibfnamefont{A.}~\bibnamefont{Lagendijk}},
  \bibinfo{journal}{Phys. Rev. B} \textbf{\bibinfo{volume}{51}},
  \bibinfo{pages}{11433} (\bibinfo{year}{1995}).

\bibitem[{\citenamefont{Averitt et~al.}(2001)\citenamefont{Averitt, Rodriguez,
  Lobad, Siders, Trugman, and Taylor}}]{AverittYBCO}
\bibinfo{author}{\bibfnamefont{R.~D.} \bibnamefont{Averitt}},
  \bibinfo{author}{\bibfnamefont{G.}~\bibnamefont{Rodriguez}},
  \bibinfo{author}{\bibfnamefont{A.~I.} \bibnamefont{Lobad}},
  \bibinfo{author}{\bibfnamefont{J.~L.~W.} \bibnamefont{Siders}},
  \bibinfo{author}{\bibfnamefont{S.~A.} \bibnamefont{Trugman}},
  \bibnamefont{and} \bibinfo{author}{\bibfnamefont{A.~J.}
  \bibnamefont{Taylor}}, \bibinfo{journal}{Phys. Rev. B}
  \textbf{\bibinfo{volume}{63}}, \bibinfo{pages}{140502}
  (\bibinfo{year}{2001}).

\bibitem[{\citenamefont{Kaindl et~al.}(2000)\citenamefont{Kaindl, Woerner,
  Elsaesser, Smith, Ryan, Farnan, McCurry, and Walmsley}}]{KaindlRA:UltmrY}
\bibinfo{author}{\bibfnamefont{R.~A.} \bibnamefont{Kaindl}},
  \bibinfo{author}{\bibfnamefont{M.}~\bibnamefont{Woerner}},
  \bibinfo{author}{\bibfnamefont{T.}~\bibnamefont{Elsaesser}},
  \bibinfo{author}{\bibfnamefont{D.~C.} \bibnamefont{Smith}},
  \bibinfo{author}{\bibfnamefont{J.~F.} \bibnamefont{Ryan}},
  \bibinfo{author}{\bibfnamefont{G.~A.} \bibnamefont{Farnan}},
  \bibinfo{author}{\bibfnamefont{M.~P.} \bibnamefont{McCurry}},
  \bibnamefont{and} \bibinfo{author}{\bibfnamefont{D.~G.}
  \bibnamefont{Walmsley}}, \bibinfo{journal}{Science}
  \textbf{\bibinfo{volume}{287}}, \bibinfo{pages}{470 } (\bibinfo{year}{2000}).

\bibitem[{\citenamefont{Kabanov et~al.}(1999)\citenamefont{Kabanov, Demsar,
  Podobnik, and Mihailovic}}]{PhysRevB.59.1497}
\bibinfo{author}{\bibfnamefont{V.~V.} \bibnamefont{Kabanov}},
  \bibinfo{author}{\bibfnamefont{J.}~\bibnamefont{Demsar}},
  \bibinfo{author}{\bibfnamefont{B.}~\bibnamefont{Podobnik}}, \bibnamefont{and}
  \bibinfo{author}{\bibfnamefont{D.}~\bibnamefont{Mihailovic}},
  \bibinfo{journal}{Phys. Rev. B} \textbf{\bibinfo{volume}{59}},
  \bibinfo{pages}{1497} (\bibinfo{year}{1999}).

\bibitem[{\citenamefont{Rothwarf and Taylor}(1967)}]{RToriginal}
\bibinfo{author}{\bibfnamefont{A.}~\bibnamefont{Rothwarf}} \bibnamefont{and}
  \bibinfo{author}{\bibfnamefont{B.~N.} \bibnamefont{Taylor}},
  \bibinfo{journal}{Phys. Rev. Lett.} \textbf{\bibinfo{volume}{19}},
  \bibinfo{pages}{27} (\bibinfo{year}{1967}).

\bibitem[{\citenamefont{Kabanov et~al.}(2005)\citenamefont{Kabanov, Demsar, and
  Mihailovic}}]{RTsolution}
\bibinfo{author}{\bibfnamefont{V.~V.} \bibnamefont{Kabanov}},
  \bibinfo{author}{\bibfnamefont{J.}~\bibnamefont{Demsar}}, \bibnamefont{and}
  \bibinfo{author}{\bibfnamefont{D.}~\bibnamefont{Mihailovic}},
  \bibinfo{journal}{Physical Review Letters} \textbf{\bibinfo{volume}{95}},
  \bibinfo{eid}{147002} (\bibinfo{year}{2005}).

\bibitem[{\citenamefont{Coleman and Schofield}(2005)}]{Coleman}
\bibinfo{author}{\bibfnamefont{P.}~\bibnamefont{Coleman}} \bibnamefont{and}
  \bibinfo{author}{\bibfnamefont{A.~J.} \bibnamefont{Schofield}},
  \bibinfo{journal}{Nature} \textbf{\bibinfo{volume}{433}},
  \bibinfo{pages}{226} (\bibinfo{year}{2005}).

\bibitem[{\citenamefont{Cox and Grewe}(1988)}]{Cox}
\bibinfo{author}{\bibfnamefont{D.~L.} \bibnamefont{Cox}} \bibnamefont{and}
  \bibinfo{author}{\bibfnamefont{N.}~\bibnamefont{Grewe}}, \bibinfo{journal}{Z.
  Phys. B} \textbf{\bibinfo{volume}{71}}, \bibinfo{pages}{321}
  (\bibinfo{year}{1988}).

\bibitem[{\citenamefont{Millis and Lee}(1987)}]{Millis}
\bibinfo{author}{\bibfnamefont{A.~J.} \bibnamefont{Millis}} \bibnamefont{and}
  \bibinfo{author}{\bibfnamefont{P.}~\bibnamefont{Lee}},
  \bibinfo{journal}{Phys. Rev. B} \textbf{\bibinfo{volume}{35}},
  \bibinfo{pages}{3394} (\bibinfo{year}{1987}).

\bibitem[{\citenamefont{Men\'endez and Cardona}(1984)}]{CardonaSiAnharm}
\bibinfo{author}{\bibfnamefont{J.}~\bibnamefont{Men\'endez}} \bibnamefont{and}
  \bibinfo{author}{\bibfnamefont{M.}~\bibnamefont{Cardona}},
  \bibinfo{journal}{Phys. Rev. B} \textbf{\bibinfo{volume}{29}},
  \bibinfo{pages}{2051} (\bibinfo{year}{1984}).

\bibitem[{\citenamefont{Balkanski et~al.}(1983)\citenamefont{Balkanski, Wallis,
  and Haro}}]{HaroAnharmSi}
\bibinfo{author}{\bibfnamefont{M.}~\bibnamefont{Balkanski}},
  \bibinfo{author}{\bibfnamefont{R.~F.} \bibnamefont{Wallis}},
  \bibnamefont{and} \bibinfo{author}{\bibfnamefont{E.}~\bibnamefont{Haro}},
  \bibinfo{journal}{Phys. Rev. B} \textbf{\bibinfo{volume}{28}},
  \bibinfo{pages}{1928} (\bibinfo{year}{1983}).

\bibitem[{\citenamefont{Hart et~al.}(1970)\citenamefont{Hart, Aggarwal, and
  Lax}}]{LaxAnharmSi}
\bibinfo{author}{\bibfnamefont{T.~R.} \bibnamefont{Hart}},
  \bibinfo{author}{\bibfnamefont{R.~L.} \bibnamefont{Aggarwal}},
  \bibnamefont{and} \bibinfo{author}{\bibfnamefont{B.}~\bibnamefont{Lax}},
  \bibinfo{journal}{Phys. Rev. B} \textbf{\bibinfo{volume}{1}},
  \bibinfo{pages}{638} (\bibinfo{year}{1970}).

\bibitem[{\citenamefont{Matsushita}(1976)}]{MATSUSHITAM:Anotdt}
\bibinfo{author}{\bibfnamefont{M.}~\bibnamefont{Matsushita}},
  \bibinfo{journal}{J. Chem. Phys.} \textbf{\bibinfo{volume}{65}},
  \bibinfo{pages}{23 } (\bibinfo{year}{1976}).

\bibitem[{\citenamefont{Lockwood et~al.}(1983)\citenamefont{Lockwood, Katiyar,
  and So}}]{PhysRevB.28.1983}
\bibinfo{author}{\bibfnamefont{D.~J.} \bibnamefont{Lockwood}},
  \bibinfo{author}{\bibfnamefont{R.~S.} \bibnamefont{Katiyar}},
  \bibnamefont{and} \bibinfo{author}{\bibfnamefont{V.~C.~Y.} \bibnamefont{So}},
  \bibinfo{journal}{Phys. Rev. B} \textbf{\bibinfo{volume}{28}},
  \bibinfo{pages}{1983} (\bibinfo{year}{1983}).

\bibitem[{\citenamefont{Wakamura}(1989)}]{WAKAMURAK:Obssfe}
\bibinfo{author}{\bibfnamefont{K.}~\bibnamefont{Wakamura}},
  \bibinfo{journal}{Sol. St. Comm.} \textbf{\bibinfo{volume}{71}},
  \bibinfo{pages}{1033 } (\bibinfo{year}{1989}).

\bibitem[{\citenamefont{Doniach}(1977)}]{Doniach_kondalattice}
\bibinfo{author}{\bibfnamefont{S.}~\bibnamefont{Doniach}},
  \bibinfo{journal}{Physica B + C} \textbf{\bibinfo{volume}{91B+C}},
  \bibinfo{pages}{231 } (\bibinfo{year}{1977}).

\end{thebibliography}
\end{document}